\documentclass[sigplan,nonacm]{acmart}
\startPage{1}

\setcopyright{none}

\usepackage{booktabs}   
\usepackage{subcaption} 
\usepackage{xcolor}
\usepackage{url}
\usepackage{amsmath}
\usepackage{mathtools}
\usepackage{listings}

\lstdefinelanguage{CollectionsP}{%
  language     = Python,
  morekeywords = {pclass, Q, I, P, from, join, select, on, where,
    union, intersect, select, or, and, is, with, as, foreach,
    subset, union, intersect, Join, Project, set, subset, projection, parameter, Getter, GetterSetter, Deleter, Setter, Producer},
}
\begin{document}

\title[GoTcha]{GoTcha: An Interactive Debugger for GoT-Based Distributed Systems}         


\author{Rohan Achar}
\affiliation{
  \department{Donald Bren School of ICS}              
  \institution{University of California, Irvine}            
  \city{Irvine}
  \state{CA}
  \postcode{92697}
  \country{USA}                    
}
\email{rachar@ics.uci.edu}          

\author{Pritha Dawn}
\affiliation{
  \department{Donald Bren School of ICS}              
  \institution{University of California, Irvine}            
  \city{Irvine}
  \state{CA}
  \postcode{92697}
  \country{USA}                    
}
\email{pdawn@ics.uci.edu}          

\author{Cristina V. Lopes}
\affiliation{
  \department{Donald Bren School of ICS}              
  \institution{University of California, Irvine}            
  \city{Irvine}
  \state{CA}
  \postcode{92697}
  \country{USA}                    
}
\email{lopes@ics.uci.edu}          

\begin{abstract}
Debugging distributed systems is hard. Most of the techniques that have been developed for debugging such systems use either extensive model checking, or postmortem analysis of logs and traces. Interactive debugging is typically a tool that is only effective in single threaded and single process applications, and is rarely applied to distributed systems. While the live observation of state changes using interactive debuggers is effective, it comes with a host of problems in distributed scenarios. In this paper, we discuss the requirements an interactive debugger for distributed systems should meet, the role the underlying distributed model plays in facilitating the debugger, and the implementation of our interactive debugger: GoTcha.

GoTcha is a browser based interactive debugger for distributed systems built on the Global Object Tracker (GoT) programming model. We show how the GoT model facilitates the debugger, and the features that the debugger can offer. We also demonstrate a typical debugging workflow.
\end{abstract}


\maketitle

\begin{abstract}
Debugging distributed systems is hard. Most of the techniques that have been developed for debugging such systems use either extensive model checking, or postmortem analysis of logs and traces. Interactive debugging is typically a tool that is only effective in single threaded and single process applications, and is rarely applied to distributed systems. While the live observation of state changes using interactive debuggers is effective, it comes with a host of problems in distributed scenarios. In this paper, we discuss the requirements an interactive debugger for distributed systems should meet, the role the underlying distributed model plays in facilitating the debugger, and the implementation of our interactive debugger: GoTcha.

GoTcha is a browser based interactive debugger for distributed systems built on the Global Object Tracker (GoT) programming model. We show how the GoT model facilitates the debugger, and the features that the debugger can offer. We also demonstrate a typical debugging workflow.
\end{abstract}
\section{Introduction}
\label{sec:intro}

Debugging faults and errors in distributed systems is hard. There are many reasons for the inherent difficulty, and these have been extensively discussed in the literature~\cite{testing1,shiviz,valadares2016}. In particular, the non determinism of concurrent computation and communication makes it hard to reliably observe error conditions and failures that expose the bugs in the code. The act of debugging, itself, may change the conditions in the system to hide real errors or expose unrealistic ones.

To mitigate the effects of these difficulties, several approaches have been developed. These approaches range from writing simple but usually inadequate test cases~\cite{testing2}, to rigorous but expensive model checking~\cite{modelchecking1} and theorem proving~\cite{verdi}. More recently, techniques such as record and replay~\cite{recordreplay}, tracing~\cite{tracing}, log analysis~\cite{loganalysis}, and visualizations~\cite{shiviz} have been used to locate bugs by performing postmortem analysis on the execution of distributed systems after errors are encountered.

None of these tools are interactive debuggers. Interactive debuggers in a single threaded application context are powerful tools that can be used to pause the application at a predetermined predicate (often a line of code in the form of a breakpoint) and observe the state of the variables and the system at that point. They can observe the errors as they happen, and can be quite effective in determining the cause. Controls are often provided to execute each line of code interactively and observe the change of state after each step. Many modern breakpoint debuggers provide the ability to roll back the execution to a line that was already executed. This, along with the ability to mutate the state of the system from the debugger, can be used to execute the same set of lines over again and observe the state changes without having to restart the application.


\begin{figure}
\centering
\includegraphics[width=0.4\textwidth]{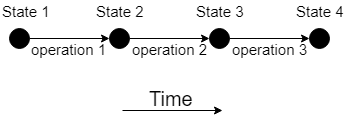}
\caption{Information propagation in single-thread systems.}
\label{fig:single_thread}
\end{figure}

\begin{figure}
\centering
\includegraphics[width=0.45\textwidth]{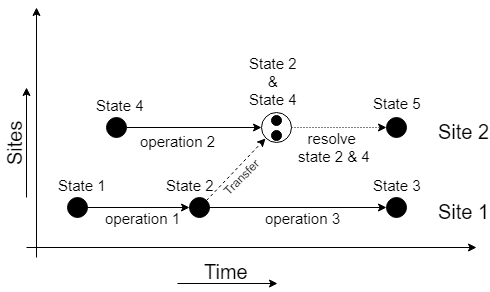}
\caption{Information propagation in distributed systems.}
\label{fig:distributed}
\end{figure}

Traditional interactive debuggers, however, become inadequate when used in parallel or distributed systems. Techniques used in single threaded applications do not translate well to a parallel or distributed context because information creation and consumption is not sequential. To create an interactive debugger for a distributed context, information flow must be modeled differently.

This paper dives into the topic of interactive debuggers for distributed systems and presents an approach to the problem based on a specific model of distributed programming recently proposed, GoT~\cite{got}. The paper is organized as follows. First, we discuss the problem and existing approaches to debugging distributed systems in Section~\ref{sec:related}. In Section~\ref{sec:design}, we analyze the requirements to design an effective interactive debugger for distributed systems. This includes both the features that the debugger should have, and how we would go about designing those features. The underlying distributed computing model plays a dominant role in determining the feasibility of interactive debugging and so, in Section~\ref{sec:ideal_model}, we determine the essential features that a distributed computing model must have in order to create an effective interactive debugger. We put our approach to the test with an implementation of a distributed debugger called GoTcha. GoTcha is build on top of the GoT distributed programming model which is discussed in Section~\ref{sec:spacetime}. The implementation of GoTcha using the example of a distributed word frequency counting application, is discussed in Section~\ref{sec:debug_arch}. We show how GoTcha meets the requirements of an interactive debugger in Section~\ref{sec:meeting_req} and discuss what might be in the future for interactive debugging in distributed computing in Section~\ref{sec:discussion}.
\section{Debugging Distributed Systems}
\label{sec:related}

Interactive debugging of parallel and distributed systems has been discussed as early as 1981~\cite{schiffenbauer1981interactive}, but the idea has never been fully realized, mainly because it is very hard to do. However, there are many non interactive tools aid developers in debugging distributed applications. A comprehensive survey of the types of methods available can be found in Beschastnikh et al.~\cite{shiviz}. In this survey, existing methods are grouped into seven categories: testing, model checking, theorem proving, record and replay, tracing, log analysis, and visualization. Each of these types of tools offers different insights for the developer to find bugs in the application. Tools for record and replay~\cite{recordreplay, d3s, friday}, tracing~\cite{tracing, tracing2}, log analysis~\cite{loganalysis}, and visualizers~\cite{shiviz}, try to parse the artifacts of execution such as logs, execution stack traces, and data traces to understand the change of state in a run of the distributed system. 

Many of these tools share features with interactive debuggers, as they share the common goal of exposing errors in the system to the developers. For example, tools like ShiViz~\cite{shiviz} provide developers a way to observe the information exchanged in a distributed system by parsing logs, inferring causal relations between messages in them, and then visualizing them. Similarly, interactive debuggers for distributed systems would also need to provide a way to visualize the information being exchanged. The $\text{D}^{3}\text{S}$~\cite{d3s} tool allows programmers to define predicates that, when matched during execution, parse the execution trace to determine the source of the state changes. In interactive debugging these predicates are known as breakpoints and are the fundamental concept in interactive debugging.

Recently, a graphical, interactive debugger for distributed systems called Oddity~\cite{oddity} was presented. Using Oddity, programmers can observe communication in distributed systems. They can perturb these communications, exploring conditions of failure, reordered messages, duplicate messages etc. Oddity also supports the ability to explore several branching executions without restarting the application. Oddity highlights communication. Events are communicated, and consumed at the control of the programmer via a central Oddity component. However, the tool does not seem to capture the change of state within the node, it only captures the communication. Without exposing the change of state within the node due to these communications, the picture is incomplete. With this tool, we can observe if the wrong messages are being sent, but we cannot observe if a correct message is processed at a node in the wrong way.

\section{Fundamental Requirements of Interactive Debuggers}
\label{sec:design}
Whether for a single threaded application or a distributed application, there are two fundamental requirements that interactive debuggers must provide. First, the debugger should be able to observe, step by step, the change of state of the application. Second, the debugger must give the user control over the flow of execution, so that alternative chains of events across the distributed application components can be observed. In this section, we discuss the design choices available to us when trying to achieve both goals. 

\subsection{Requirement 1: Observing State Changes}
\label{subsec:req1}

The most important goal of an interactive debugger is to observe, step by step, the change of state of the computing system. Therefore, it is important to understand how change of state can occur. A change of state can be abstracted to the consumption and creation of information. For example, over the execution of a line of code in a single threaded application, the current state of the variables in the application's heap is consumed, and a new state is created. In such an application, we only have one dimension over which information is created and consumed: time -- not necessarily world time, but time modeled as the sequence of operations, or causal time. Figure~\ref{fig:single_thread} shows this progression. The only task that an interactive debugger designed for single threaded applications has to do is to show the change of state over the sequential execution of each line of code.

In the context of a parallel or distributed system, we have an additional dimension to think about: the site of execution (thread in parallel systems, nodes in distributed systems). Information is not only generated and consumed over lines of code at a site, it is also transmitted from one site to another and then made available at that site. Figure~\ref{fig:distributed} shows this information exchange. An interactive debugger for such a system must model three effects: the change of state over execution at each site, the transfer of state between sites, and the reconciliation of the states received from remote sites with the state at that site. It is possible to consider reconciliation as part of the state changes through the execution of code at a site. However, it is more useful, in the context of interactive debugging, to keep them separate. Reconciliation does not always occur through dedicated lines of code. Often asynchronous operations accept communications and update states. It could also just be a side effect of receiving a transmission of state. For example, when multiple clients concurrently update the same keys of the database using the last write wins reconciliation strategy, the old state is simply replaced with the new state as a part of the transfer. The overwriting of information is not implicitly recorded. Writes that are lost to this become hard to track. Interactive debuggers have a hard time highlighting these lost writes and so developers cannot use the debugger effectively when fixing related bugs. Since the point of the debugger is to enable the developer to reason over state changes and detect errors, it is better to expose reconciliation separately. 

In summary, in a distributed or parallel system, an interactive debugger needs to expose three types of state changes: changes due to local execution, transfer of state between sites, and changes due to reconciliation of multiple states at a site. We discuss each of these next.

\subsubsection{Exposing State Changes Due to Local Execution}
\label{subsubsec:local_exec}
Designing to expose state changes due to local execution is quite straightforward, and traditional interactive debuggers do it already. There is, however, an issue of scale. Since there are multiple sites to track, there are many code paths to follow. The developer can easily get overwhelmed by this. The debugger needs to filter out the unimportant parts. One way to do this would be to treat execution paths from one point of inter-site communication to the next as one unit to step through. Doing this cuts down the number of local state updates the distributed application debugger needs to follow.

\subsubsection{Exposing Transfer of State Between Sites}
There are only two ways in which state can be transferred, and every form of communication falls into one of them: pushing and pulling changes. A web browser receiving website data is pulling changes from a server. A node sending events to all other nodes that have subscribed to the events is pushing data. Since these are the only two ways in which state can be transferred, the debugger must pay special attention to these two primitives in any distributed model and expose the calls to these primitives explicitly to the developer. 

Pull and push operations consist of one or more phases. At a minimum, the sequence for a pull operation includes a request for information, and then information is received in response to the request; similarly, the minimum for a push operation is one command wherein the information is sent. More robust implementations, however, have multiple phases with acknowledgements. Several distributed models optimize by making these calls asynchronous and sometimes going to the extent of taking the control over communication away from the programmer. For example, in the publish-subscribe model, the subscriber of data receives data via a push operation when the data is published at a different site. 

\subsubsection{Exposing Changes Due to Reconciliation of Multiple States}
\label{subsubsec:recon}
When a node receives state changes from another site, it needs to reconcile the state. It is important to understand reconciliation, and differentiate it from conflict resolution. 

Reconciliation is a two step process. The first step is to receive the information of state change from another site. The second step is conflict resolution where the information received is meaningfully merged with the information already present in the site, to make the local state coherent for the next local execution. Different distributed models deal with these two steps in different ways, making them particularly tricky to observe. 

In most distributed models, the two steps happen together. Remote changes are evaluated as soon as they are received and decision is taken regarding their incorporation into the local state, e.g. last-write-wins, CRDTs~\cite{shapiro11}. In these models, observing conflict resolution is the same as observing reconciliation. In some distributed models, however, state changes received are stored, and conflict resolution, if any, is deferred to a later time. For example, total store ordering~\cite{tso}, global sequence protocols~\cite{gsp, gsp1}, TARDiS~\cite{tardis}, Irmin~\cite{mergeable_types}, concurrent revisions~\cite{semantics-of-concurrent-revisions-2}, GoT~\cite{got}, are a few models that first store the incoming changes, and provide the programmer control over when these changes are resolved and introduced into the local state. From the point of view of the user of an interactive debugger observing reconciliation in these models, the user must both observe when information is received from a remote site, and when the information is accepted and incorporated in the local state.

Looking at conflict resolution in particular, there are a myriad of ways in which concurrent state updates are resolved, and it entirely depends on the underlying distributed model. Some models such as the last-write-wins, total store ordering~\cite{tso}, global sequence protocols~\cite{gsp, gsp1}, etc. resolve conflicts implicitly. Since many of the models do not retain causal relations between reads and writes of state, it is hard to tell if an overwrite was an intended update, or the result of implicit conflict resolution. As such, it becomes quite difficult for an interactive debugger to expose the point of conflict resolution in such models. Other models such as TARDiS, Irmin, concurrent revisions, and GoT resolve conflicts explicitly using programmer-written merge functions. Although in many models these merge functions are called asynchronously, there exist specific execution paths which deal with conflict resolution, and this can be exposed by the interactive debugger.


\subsection{Requirement 2: Controlling the Flow of Execution}
\label{subsec:req2}

Interactive debuggers, as the name suggest, must allow the user to debug the distributed application interactively. To be interactive, the debugger must take control of all forms of state changes present in the distributed system and hand this control over to the user. In a single threaded system, with only one form of state change and executed at a single location, taking control of the execution and handing it over to the user is relatively straightforward. However, in a distributed system, this is harder. There are more forms of state changes as described above, and these state changes can execute over multiple sites. If the interactive debugger is controlled by the user on only one site, and the rest of the sites are free to execute, then the user has control over only one form of state change: changes occurring due to local execution.

In order for the user to have control and observe all forms of state change, the user needs to be able to pause all sites when one site is paused. The problems associated with distributed computing are inherited by the debugger trying to exercise global control over the system. An easy solution, one that is present in traditional interactive debuggers when trying to debug multi-threaded programs, is to pause all threads of execution when one thread is paused. For example, the GDB debugger has an all-stop mode\footnote{https://sourceware.org/gdb/onlinedocs/gdb/Thread-Stops.html} that behaves in that manner. While this solution can work in simple multi-threaded applications operating out of the same machine, this approach, used as is, becomes difficult when moving to the distributed context when sites are located at different machines. Triggering a pause on one machine would have to be made instantly visible to all nodes, which is hard, as there are inevitable network delays, leading to unintended state changes after the user has tried to exert control.

The problem of exerting global control is even more pronounced when each site communicates with multiple sites, such as in peer-to-peer applications. In a server-client model where all clients only communicate with the server, pausing the server could allow us to pause the clients. A solution then, perhaps, could be to transform the distributed system into a server-client model with an interactive debugger as the central component. All forms of state changes could be rerouted through this central component, giving this component the ability to observe and control all these forms of state change.

Rerouting both networked and local state changes would significantly alter the network conditions for the system. This is fine, as long as the user of the system is able to leverage the interactivity gained to explore state changes in the application related to different orderings of concurrent operations. The advantage offered depends on the system being developed. If there are too many variations or ordering possible (e.g. a large system with many sites), the developer might not be able to observe them all.

While exploring the design choices available to us when fulfilling these two requirements, it becomes clear that the underlying distributed model on which the interactive debugger is to be built on is absolutely dominant. The communication, and state reconciliation methods used by the model play a heavy role in determining the capabilities that the interactive debugger has. That said, in the next section, we explore what support from the distributed model is necessary to make an interactive debugger viable.
\section{Constraints on Distributed Systems}
\label{sec:ideal_model}

There are a large number of distributed computing models. Each model optimizes for different goals and, therefore, makes different design choices for different aspects under its control. Some of the choices can help an interactive debugger expose the state changes of the system accurately to the developer, while others can hinder it. If interactive debugging is a goal, the underlying distributed system model must be constrained in specific ways. In this section, we discuss at least three constraints that a distributed model must abide in order to support the requirements of interactive debuggers discussed in the previous section.

\subsection{Read Stability}
As discussed in the previous section, there are three ways in which information is created and consumed in a distributed context: state change through local execution, transfer of state between nodes, and state change through reconciliation of multiple states at a node. The distributed model must expose these three ways as separate events on their own. An important constraint that the model must have in order to be able to separate these three scenarios is read stability.

Read stability -- a concept typically associated with isolation in database transactions~\cite{readstability} -- is a property of the model where changes to the local state can only occur when the local execution context wants it to. There are two ways to affect change to the local state: writes from local execution, and reconciliation of local state and a state that is received from an external node. Read stability can be easily achieved if the model, after the transfer of information between sites, does not reconcile the information immediately but instead stores the information (cache, queue, etc.) and waits for the local execution to accept these changes. For example, a multi-threaded system with multiple threads writing concurrently to shared variables, without locks, breaks the read stability requirement.

Without read stability, the interactivity of the debugger is heavily curtailed. For example, when the user of an interactive debugger debugging an application paused at a certain execution point, steps through one line of execution, the expectation from the user is that the state change being observed is a state change due to the execution of that one line. However, in a system without read stability, this might not be true. The local state change could also have changed via reconciliation with external changes, essentially giving rise to a situation where there are multiple types of state change being executed in the same step. Interactive debugging is a debugging method that aims to give the control of execution to the user with the aim of letting the user observe all forms of state change explicitly. Therefore, having multiple types of state change being executed in a single step is not acceptable. Read stability solves this problem by ensuring that updates to the local state are explicitly defined and occur only when the local execution context expects it.

There are many methods to achieve read stability. For example, in the global sequence protocol (GSP) model~\cite{gsp1, gsp}, the operations that are distributed to all remote sites are placed in an pending queue, ready to be applied through an explicit primitive given to the local application. In TARDiS~\cite{tardis}, concurrent writes are placed under version control in separate branches to avoid interfering with each other until one context wants to introduce these concurrent changes to its branch using the resolve primitive. 


\subsection{Separation of Published State and Local State}
When a local site makes changes locally, some models immediately make these changes available for other sites to observe. This means that the local state of a site at the end of every line of code is potentially a site that is going to be observed by a remote node, which also means that each of these states have to be tracked by the debugger. Observing state change over local execution over every line of code in all sites of a distributed system is definitely not scalable and the information can be overwhelming to the user observing it.

Some models (for e.g. publish-subscribe), do not make changes, made locally, immediately available for other sites. These changes are made ready for consumption only when they are specifically published. The local state of the site is kept separate from what the site wants to share. An interactive debugger for the system just needs to take control over the points where the site shares information, grouping all changes to local state in between as a single operation. This makes the observable set of operations smaller where we trade in the fine granularity of observing change of state over every line of code to look at the execution in broader strokes which makes the implementation feasible. Therefore, the separation of published state and local state is critical when attempting to create an interactive debugger at scale.

\subsection{Explicit Mechanisms for State Change}
Just as the local execution context should have control over the introduction of changes to its state, the local execution should also have control over when the local changes are being transmitted to other nodes. Having an explicit primitive to start the transfer of changes helps the debugger expose the start point of transfer. Having explicit primitives deal with receiving these changes, and then introduce these changes to the local state helps the debugger expose reconciliation between the transferred state and the local state.

\section{The GoT Distributed Computing Model}
\label{sec:spacetime}
As mentioned before, the underlying distributed computing model has a dominant influence on the feasibility of an interactive debugger. Our interactive debugger, GoTcha, is designed for a specific distributed programming model called Global Object Tracker (GoT)~\cite{got}. Specifically, GoTcha is built as a debugger for a Python implementation of GoT called Spacetime. In this section, we describe GoT, and explain its design features that are relevant to GoTcha. 

\subsection{GoT: Git, but for Objects}
A Spacetime application consists of many nodes, called GoT nodes, that perform tasks asynchronously within the distributed application. The GoT nodes share among themselves a collection of objects. Each of these nodes can be executed in different machines, communicating the changes to the state of the shared objects via the network. What is unique about GoT is that the synchronization of object state among the distributed nodes is seen as a version control problem, with a solution that is modeled after Git~\cite{got}. 

\begin{figure*}
\centering
\includegraphics[width=0.66\textwidth]{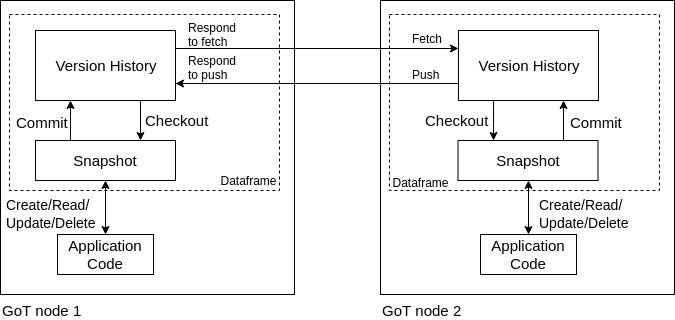}
\caption{Structure of a GoT node. Arrows denote the direction of data flow.}
\label{fig:gotnode}
\end{figure*}

All GoT nodes that are part of the same spacetime application have a repository of the shared objects, called a dataframe. The dataframe is akin to a Git repository, and, like a Git repository, it has two components: a snapshot and a version history, as shown in Figure~\ref{fig:gotnode}. The snapshot, analogous to the staging area in Git, defines the local state of the node. All changes made by the application code on the local dataframe are first staged in this snapshot. The version history, on the other hand, is the published state of the node. Like in Git, changes can be moved from the staging area to the version history using the {\em commit} primitive, and the snapshot can be made up to date with the latest version in the version history by using the {\em checkout} primitive. Inter-node communication happens using {\em push} and {\em fetch} requests, used to communicate updates in version histories between nodes. 

When the version history at a node receives changes (via {\em commit}, {\em push} or {\em fetch}), a conflict with concurrent local changes is possible and must be resolved.  While in Git conflicts are resolved manually by the user, and only on a {\em fetch}, in GoT, conflicts are resolved automatically, at the node receiving the changes, and irrespective of the primitive used. Automatic conflict resolution is achieved via programmer-defined three-way merge functions that are invoked when conflicts are detected.


The APIs supported by the dataframe are shown Table~\ref{tbl:api}; this table can be used as a quick reference for the API calls in the example explained next.

\begin{table*}
  \begin{tabular}{lll}
    \toprule
    Dataframe API & Equivalent Git API & Purpose \\
    \midrule
    read\_\{one, all\} & N/A & Read objects from local snapshot. \\
    add\_\{one, many\} & git add <untracked> & Add new objects to local snapshot. \\
    delete\_\{one, all\} & git rm <files> & Delete objects from local snapshot. \\
    \midrule
     & git add <modified> & Objects are locally modified which is tracked by the local snapshot. \\
    \midrule
    commit & git commit & Write staged changes in local snapshot to local version history. \\
    checkout & git checkout & Update local snapshot to the local version history HEAD. \\
    \midrule
    push & git push & Write changes in local version history to a remote version history. \\
    fetch & git fetch \&\& git merge & Get changes from remote version history to local version history. \\
    pull & git pull & fetch and then checkout. \\
    \bottomrule
  \end{tabular}
  \caption{API table for a dataframe}
  \label{tbl:api}
\end{table*}

\subsection{GoT Example: Distributed Word Frequency Counter}
The example that we use is a distributed word frequency counter. The application takes a file as input and shards it by line. These lines are distributed to several remotely located workers which tokenize and count the frequency of the tokens in the lines. The partial counts are then aggregated and presented by the application as the final word frequency tally. 

The distributed word frequency counter application has two types of GoT nodes: WordCounter and Grouper. The Grouper node controls the execution of this Spacetime application. It is responsible for sharding the input files into lines and aggregating partial word frequency counts to reach the final tally. The WordCounter node is responsible for tokenizing and counting the word frequencies in each line. 

\begin{figure}
\centering
\includegraphics[width=0.45\textwidth]{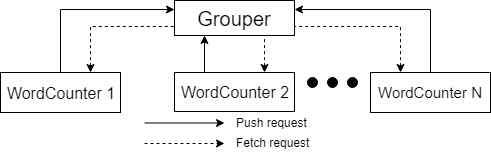}
\caption{Network topology of an example distributed word counting application built on Spacetime.}
\label{fig:network}
\end{figure}

WordCounter nodes are responsible for the communication in this Spacetime application. Every WordCounter node must {\em fetch} changes from, and {\em push} changes to the Grouper node. This relation between these nodes is shown in Figure~\ref{fig:network} and defines the network topology of our example application. 

The dataframes at each WordCounter and Grouper nodes share objects of type Line, WordCount, and Stop that are shown in Listing~\ref{lst:datamodel}. 

\begin{lstlisting}[language=Python,basicstyle=\small, numbers=left,
label=lst:datamodel, captionpos=b, caption=The types used by the Word Counting application.]
class Line(object):
    line_num = primarykey(int)
    line = dimension(str)
    def __init__(self,line_num,line):
        self.line_num = line_num
        self.line = line
    def process(self):
        # a simple tokenizer
        return self.line.strip().split()
        
class WordCount(object):
    word = primarykey(str)
    count = dimension(int)
    def __init__(self,word,count):
        self.word = word
        self.count = count

class Stop(object):
    index = primarykey(int)
    accepted = dimension(bool)
    def __init__(self, index):
        self.index = index
        self.accepted = False

\end{lstlisting}

All types registered to the dataframe have list of attributes marked as dimensions of the type, that declare the attributes to be stored and tracked in the dataframe. One dimension in each type is a special attribute and is defined as the primarykey. Objects are stored and retrieved by Spacetime by the value of this primarykey attribute. 

Objects of type Line are shards of the input file. The dimension line\_num (defined in line 2) is the primary key, of type integer, that represents the line number in the input file. The dimension, line (line 3), stores the contents of the line as a string. Objects of type WordCount store the word frequency for a unique token and have two dimensions: the primarkey, word (line 12), is a string representing the token, and count (line 13), is an integer representing the frequency of that token. WordCounter Nodes communicate completion using objects of type Stop having two dimensions: the, primarykey, index (line 19), an unique identifier representing a single WordCounter worker, and accepted (line 20), which is set to True by the WordCounter node signalling the completion of its task. Any state in attributes outside these dimensions is purely a local state, and is not tracked and shared by the dataframe.

\vspace{3cm}

\begin{lstlisting}[language=Python,basicstyle=\small, numbers=left, 
label=lst:grouper, captionpos=b, caption=The Grouper node.]
def Grouper(df,filename,num_count):
    i = 0
    for line in open(filename):
        df.add_one(Line, Line(i,line))
        df.commit(); i += 1;
    df.add_many(Stop,
      [Stop(n) for n in range(num_count)])
    df.commit()
    while not all(
          s.accepted
          for s in df.read_all(Stop)):
        df.checkout()
    for w in df.read_all(WordCount):
        print(w.word, w.count)
        
if __name__ == "__main__":
    filename, num_count = sys.argv[1:]
    grouper_node = Node(
        Grouper, server_port=8000
        Types=[Line,WordCount,Stop])
    grouper_node.start(filename,num_count)
\end{lstlisting}

Listing~\ref{lst:grouper} shows part of the application code for an instance of the Grouper node. The grouper\_node is instantiated (lines 18-20) with the application code, defined by the function Grouper, along with the types to be stored in the dataframe, and the port on which to listen to incoming connections. The grouper\_node is launched using the blocking call, start (line 21), and takes in the parameters that must be passed to this instance of the Grouper node: the input file and the number of WordCounter nodes that are going to be launched. 

The Grouper function  (line 1) that is executed receives the repository, dataframe, as the first parameter, and all run-time supplied parameters as the additional parameters. The node iterates over each line in the input file, creating new Line objects for each line. The Line objects are added to the local dataframe (line 4), similar to how new files are added to a changelist in git. After each Line object is added, these staged changes are committed to the dataframe (line 5) and are available to any remote dataframe that pulls from it. After all Line objects are added, Stop objects are added, one for each WordCounter worker in the application, and committed to the dataframe (line 6-7). Grouper now has to wait for all WordCounter workers to finish tokenizing the lines that it has published, and the state of the Stop object acts as that signal (Lines 9-12). Once every worker has accepted the Stop object associated with it, the Grouper reads all the WordCount objects in the repository and displays the word frequency to the user.

\begin{lstlisting}[language=Python,basicstyle=\small, numbers=left, 
label=lst:word_counter, captionpos=b, caption=The Word Counter node.]
def WordCounter(df,index,num_count):
    line_num=index; stop=None; line=None
    while not stop or line:
        df.pull()
        line = df.read_one(Line,line_num)
        if line:
            for word in line.process():
                # reads from the snapshot
                word_obj = df.read_one(
                    WordCount,word)
                if not word_obj:
                    word_obj = WordCount(
                        word,0)
                    df.add_one(word_obj)
                # changes only snapshot
                word_obj.count += 1
            line_num += num_count
        stop = df.read_one(Stop,index)
        # commit changes 
        # to version history
        # and push these changes
        # to remote node.
        df.commit(); df.push()
    stop.accepted = True
    df.commit()
    df.push()
if __name__ == "__main__":
    workers = []
    address = sys.argv[1]
    num_workers = int(sys.argv[2])
    for i in range(num_workers):
        wnode = Node(
            WordCounter,
            Types=[Line,WordCount,Stop],
            remote=(address, 8000))
        wnode.start_async(i, num_workers)
        workers.append(wnode)
    for w in workers:
        w.join()
\end{lstlisting}

Listing~\ref{lst:word_counter} shows part of the code for an instance of the WordCounter. Multiple instances of the WordCounter node are instantiated with the remote address of the Grouper node, and the same Types that Grouper uses (lines 32-35). Each instance is started asynchronously with the parameters that it needs (line 36). 

The application code for WordCounter (function WordCounter shown in lines 1-26) also takes the dataframe as the first parameter. An independent and new dataframe is created for each instance of WordCounter, and they all have the same Grouper node as the remote node. The WordCounter keeps pulling changes from the remote node (line 4) for as long as there is a new line to read in the updated local dataframe and until a Stop object associated with the instance is read in the local dataframe. In each pull cycle, the WordCounter reads a Line object from the local dataframe, using index (line 5), and tokenizes it (line 7). For every word in the token list, the node retrieves the WordCount object associated with the word from the dataframe (line 9), creating and new object if it does not exist (line 11-14), and increments the count dimension in the object by one (line 16). These updates (both new objects, and updates to existing objects), staged in the local snapshot, are committed to the local dataframe and pushed to the remote Grouper node (line 23). The WordCounter ends operations if after pulling updates from Grouper, a Stop object is present in the dataframe, and there are no new Line objects to read. The stop object is accepted by setting the accepted dimension to True and this update is committed and pushed to Grouper as the last operations by the WordCounter node.

\subsection{Dataframe: Object Repository}
To the code in each of the GoT nodes, the dataframe acts an object heap that consists of in-memory objects that are under version control. As explained above, the dataframe in each node has two components: a snapshot and an version history.

The version history is stored as a directed acyclic graph where each vertex of the graph represents one version of the state and has a globally unique identifier that labels it and each directed edge of the graph represents a causal ``happened-after'' relation. Each edge is associated with a delta of state changes (diff) that when applied to the state of the objects at the source version transform it to the state of the objects at the destination version. The latest version of the node state (called the HEAD) is the state of the node that is observable to the other nodes in the application. Changes that are staged in the snapshot cannot be observed by external nodes until they are put into the version history. 

An edge with an associated delta is added into the graph for each new version of the state and, therefore, memory usage of the application can potentially be high. To manage the memory, Spacetime implements an effective garbage collector in each node that cleans up obsolete versions.

Changes that are made to the objects in the application code, are staged in the snapshot. When the {\em commit} primitive is invoked, a new version is created in the version history, representing the newly committed state. An edge is added from the last version the snapshot was in, to the newly created version in the version history. The diff that was committed is then associated with this edge. Changes to the version history, updated by external nodes, are introduced to the snapshot via the {\em checkout} primitive.

Inter-node communication only happens between the version histories of the corresponding nodes. As seen in the example and like Git, nodes communicate changes between version histories using the {\em fetch} and {\em push} primitives present in the dataframe. {\em Fetch} retrieves changes published to the version history in a remote GoT node and applies the changes to the local version history. {\em Push} takes changes published to the local version history and delivers it to a remote GoT node. GoT takes advantage of the diffs stored in the version histories to communicate via delta encoding, reducing the amount of data transferred between nodes. 

\subsection{Conflict Detection and Resolution}
Conflicts are detected (at any node that is receiving data), when an update received is a change from a version that is not the HEAD version in the local version history. Intuitively, this means that at least two different nodes committed concurrent changes after having read the exact same version. When conflicts are detected, they are resolved using a user defined three way merge function that includes the state that was read (the original), the changes already in the version history (yours) and the conflicting changes that are incoming (theirs). The output of the merge function, much like the merge resolution in git, adds a new merged version to the version history that has a happened-after relation with both the diverging versions. 

In the WordCounter example, the state of the WordCount objects created and updated by different WordCounter nodes can be in conflict with each other when changes are pushed to the Grouper node. For example, if two WordCounter nodes concurrently read the same word in two different lines and the word has not been observed before, both the nodes would create a new WordCount object for the word. When both changes are pushed to the Grouper node, a conflict is detected and a merge function is called.

\begin{lstlisting}[language=Python,basicstyle=\small, numbers=left, 
label=lst:merge, captionpos=b, caption=Merge function used at the Grouper node.]
def merge(conf_iter,orig_df,
          your_df,their_df):
    your_df.update_not_conflicting(
        their_df)
    for orig,yours,theirs in conf_iter:
        assert isinstance(yours,WordCounter)
        yours.count += theirs.count
    return your_df
...
    # Updated Node initialization
    grouper_node = Node(
        Grouper,server_port=8000,
        Types=[Line,WordCount,Stop],
        conflict_resolver=merge)
\end{lstlisting}


An example merge function is shown in Listing~\ref{lst:merge}. This function is called asynchronously when a conflict is detected, and is used to only to reconcile conflicting state updates. The merge function receives four parameters: an iterator of all objects that have direct contradictory changes that cannot be auto resolved ({\em conf\_iter}), as well as three snapshots of the state, one for the point where the computation forked ({\em orig\_df}), and two for the version at end of the conflicting paths ({\em your\_df}, {\em their\_df}). 

In the merge function shown, objects (Line, WordCount, and Stop objects) that are new or modified in the incoming change but do not have conflicting changes in the local history are first accepted (line 3). For the objects that are in conflict (only WordCount objects can be in conflict), we read the states at three versions of the objects: the state at the fork version, and the two states at the conflicting versions -- {\em orig}, {\em yours}, and {\em theirs} from the iterator (line 3), respectively. Then, the dimension count in objects that have been updated together are added up and stored in the object tracked by the {\em your\_df} snapshot. At the end, this modified version of {\em your\_df} is considered to be the reconciled state and returned to the version history. 

There is a bug in this merge function as it does not add counts correctly. We will use this bug to demonstrate the capabilities of the interactive debugger. For quick reference, the correct merge function is shown in Listing~\ref{lst:correct_merge} at the end of Section~\ref{sec:debug_arch}.

\subsection{GoT: Enabling an Interactive Debugger}
\label{sec:meeting_constraints}

As discussed in Section~\ref{sec:ideal_model}, the distributed model needs to have the following properties to make an interactive debugger feasible: first, the nodes need to have read stability; second, the published state at each node must be separate from the local state used by the application code at that node; finally, the model must expose primitives for the transfer and reconciliation of states between nodes. We explain how GoT incorporates these properties in its design.

{\bf Read Stability}: Each GoT node computes only on the objects in the snapshot. The snapshot can only be updated with external changes when the {\em checkout} or pull primitives are invoked. Since these are invoked by the application code at the node, and not automatically behind the scenes, the snapshot does not change unless it is directed to by the application code. Therefore, GoT supports read stability.

{\bf Separation of published state and local state}: GoT also separates the published state (the version history) from the local state (the snapshot). All changes received via a {\em fetch} or {\em push}  request will only include changes that have been committed by the sender node. 

{\bf Explicit mechanisms for communication and reconciliation}: The dataframe has an explicit mechanism for inter-node communication ({\em fetch}, {\em push} ) and conflict resolution (merge functions) that can be tracked and used by the debugger to observe both the transfer of state and the reconciliation of state updates between nodes.

\section{GoTcha}
\label{sec:debug_arch}

\begin{figure*}
\centering
\includegraphics[width=0.85\textwidth]{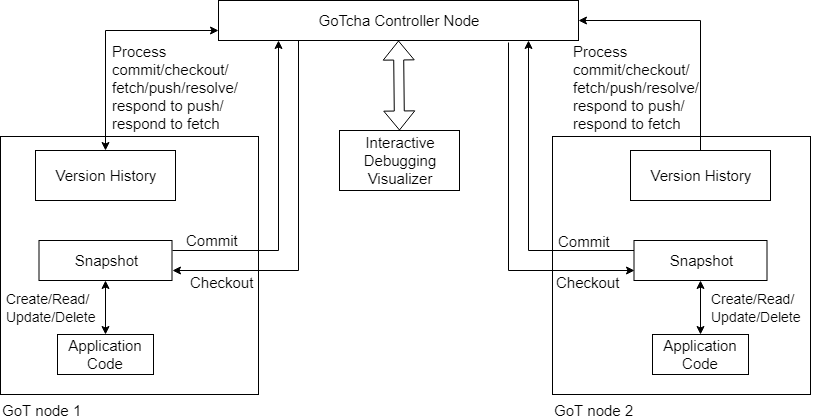}
\caption{Architecture of GoTcha.}
\label{fig:proto_arch}
\end{figure*}

GoT nodes execute their tasks over shared objects that are stored in the version history in the dataframe. These version histories are primarily used, in GoT, to facilitate the communication between nodes using delta encoding and to detect and resolve conflicts. The version history is an internal component of the dataframe and is, therefore, typically not exposed to the programmer. 

In version control systems, the version histories are more than just a datastores for files. They document the evolution of the files stored in the repository over time. There are many tools available, such as GitKraken~\footnote{https://www.gitkraken.com} and SourceTree~\footnote{https://www.sourcetreeapp.com/}, that expose this evolution to users. Observing the version history, through these tools, not only tells us the current state of the repository, but also all the changes that were made to the repository in the past and in the order that they were made. This same principle can be leveraged in GoT, to expose the version history of object changes to the user. A debugger for GoT can expose the version history allowing users to observe the evolution of the state at a node, and detect errors that have already occurred. In addition to viewing errors in the version history, live and interactive debugging becomes possible, as the updates to the version history is driven explicitly by the application code, and performed via a small API in the dataframe (see Table~\ref{tbl:api}). By taking control of these APIs and giving this control of the execution to the user, the user can stop the application, observe the state of the version history at each node, resume and observe the change of state over the execution of the dataframe operations. This, along with the ability to observe variations in the order of execution, will assist the user in observing errors as they occur. 

We created an interactive debugger called GoTcha, to expose the changes made to the version history at each node in a Spacetime application. In this section, we explain the features of GoTcha. We will continue to use the distributed word frequency counter example, detailed in the previous section, to showcase the features of the debugger.

For the purpose of the example, a test input file was created consisting of six lines, each with one word -- see Listing~\ref{lst:test_file}. The word frequencies for the words foo, bar, and baz are one, four, and one respectively. The application consists of two WordCounter nodes and one Grouper node that are launched in different machines. During execution, as shown in Listing~\ref{lst:grouper}, the Grouper node adds six Line objects and two Stop objects into its dataframe, and waits for the Stop objects to be accepted by the WordCounter nodes (Listing~\ref{lst:grouper}). WordCounter1 reads, tokenizes, and counts words on lines 1, 3, and 5. WordCounter2 does the same for lines 2, 4, and 6. Finally, both WordCounter nodes accept their Stop objects, and execution completes. The expected output is shown in Listing~\ref{lst:expected}. However, a different output is observed, shown in Listing~\ref{lst:observed}. The observed output is wrong, and this is where GoTcha can help.

\begin{figure*}
  \begin{subfigure}[b]{0.3\linewidth}
        \begin{lstlisting}[language=Python,basicstyle=\small,
        label=lst:test_file, captionpos=b, caption=Input file.]
        foo
        bar
        bar
        baz
        bar
        bar
        \end{lstlisting}
  \end{subfigure}
  \begin{subfigure}[b]{.3\linewidth}
        \begin{lstlisting}[language=Python,basicstyle=\small,
        label=lst:expected, captionpos=b, caption=Expected output.]
        foo 1
        bar 4
        baz 1
        \end{lstlisting}
  \end{subfigure}
  \begin{subfigure}[b]{.3\linewidth}
        \begin{lstlisting}[language=Python,basicstyle=\small,
        label=lst:observed, captionpos=b, caption=Observed output.]
        foo 1
        bar 6
        baz 1
        \end{lstlisting}
  \end{subfigure}
\end{figure*}

\subsection{Operation}
GoTcha follows the centralized service approach, discussed in Section~\ref{sec:design}, to have complete control over the nodes and expose the version history to the user at each node. This central service is an application by itself and is launched before any application nodes are launched. We call this application the GoTcha Controller Node (GCN from now on). The GCN is a web service to both the User Interface (UI), and the nodes in the application. When GoT nodes are launched in debug mode, they register themselves with the GCN. 

When in debug mode, the architecture of the application is modified from what is shown in Figure~\ref{fig:gotnode} to what is shown in Figure~\ref{fig:proto_arch}. The primitives that read or write from the version history at each GoT node ({\em commit}, {\em checkout}, {\em fetch}, {\em push}) are all rerouted through the GCN. With each interaction, the version history at each node is also sent to the GCN to be shown to the users. While a traditional interactive debugger for a single threaded application would observe the change of state between each line of code, GoTcha observes state changes over each action of read or write performed on the version history at each node.

\begin{figure*}
\centering
\includegraphics[width=\textwidth]{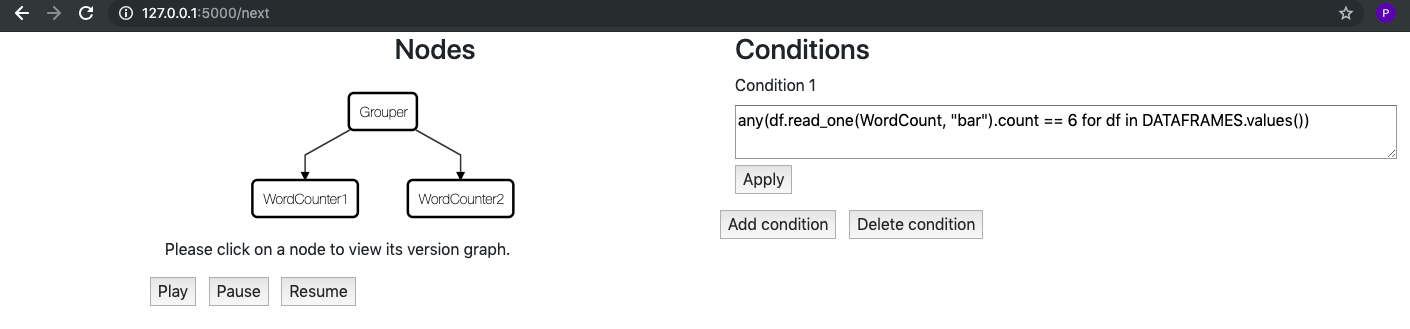}
\caption{Debugger showing the network topology of the application.}
\label{fig:start_screen}
\end{figure*}

At the start of the debugging session, after every node has been launched in debug mode, the user is shown the network topology of the application, as shown in Figure~\ref{fig:start_screen}. In this figure, on the left, the user sees that two WordCounter nodes (WordCounter1 and WordCounter2) and one Grouper node are being controlled by the GCN. The Grouper node is the authoritative node in the application, with both the WordCounter nodes making {\em fetch} and {\em push}  requests to the Grouper node. On the right, there is an input field for the user to add one or many breakpoint conditions to the debugger. The breakpoint condition shown here, returns True if there exists any WordCount object with the count dimension set to six, in the dataframe, at any GoT node.

\subsection{Observing Node State}

\begin{figure*}
\centering
\includegraphics[width=\textwidth]{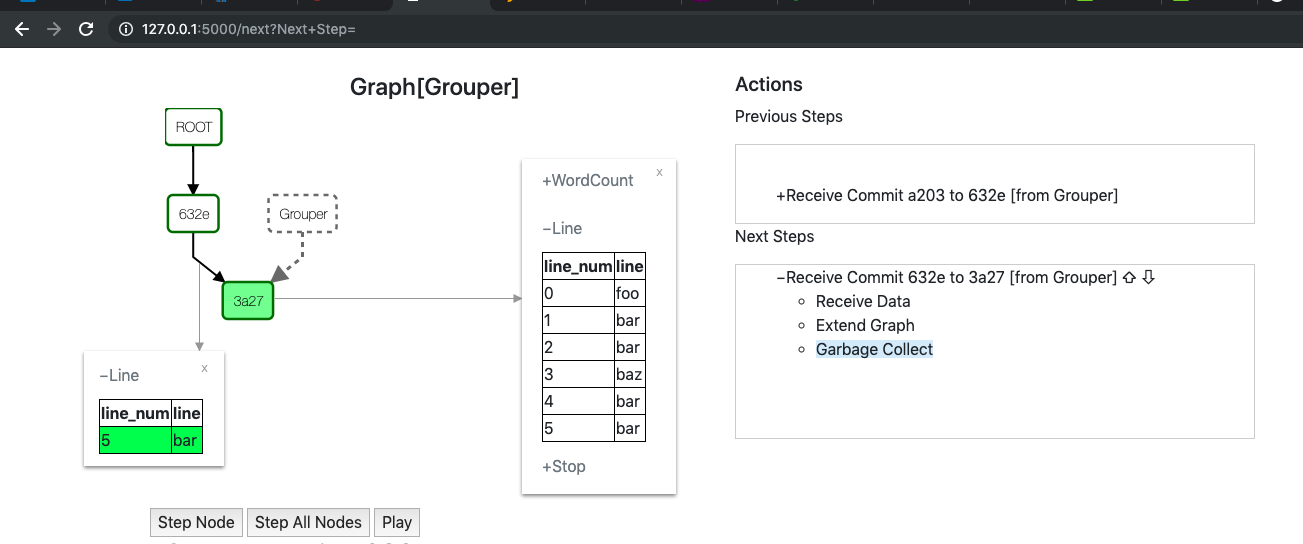}
\caption{Debugger view showing version history at the end of a {\em commit}.}
\label{fig:after_commit}
\end{figure*}

The current version history of any node can be observed by clicking on the node in the topology graph in Figure~\ref{fig:start_screen}. Figure~\ref{fig:after_commit} shows the node view of the Grouper node, observing the result of the execution of the {\em commit} primitive at line 5 of Listing~\ref{lst:grouper} (during the last iteration of that loop). The version history is shown on the top left. The history shows three versions: ROOT, 632e, and 3a27. 3a27 happened-after 632e which in turn happened-after ROOT (the start version of every version history). The HEAD version of the graph, highlighted in green, is 3a27. 

Selecting the version brings up a tooltip that shows in tabular form, the state of objects at that version. The tooltip shows that the state at version 3a27 has six objects of type Line, and shows the values of the two dimensions (line\_num, line) for the Line objects. Though the tables for Stop and WordCounter have been collapsed, as shown by the plus shaped user interface element, there are no objects of those types present yet.

Selecting the edge brings up a tooltip that shows the delta change (diff) associated with that edge. The diff associated with the edge 632e $\rightarrow$ 3a27 is also shown on the bottom left. In this case, the diff consists of a single object of type Line with the dimensions line\_num, and line having the values 5, and `bar' respectively. The entry is also marked in green, which signifies that the entry is a newly added object (added in line 4, Listing~\ref{lst:grouper}). Uncolored entries are considered to be modifications, and entries marked in red are considered to be deleted objects. The state of every version, and the diff associated with every edge can be observed. The dotted line relation shows us that the state of the snapshot of the Grouper node is known to be at version 3a27. 

On the right of Figure~\ref{fig:after_commit}, we see the state of the actions being executed on the dataframe at the Grouper node. The user sees both a list of previous steps that have been executed on the version history, and a list of steps that have to be executed (next steps). At the top of the next steps list is the current active step being executed. Each step directly maps to one of the dataframe primitives rerouted through GoTcha and is broken up into several phases. 

We can see that the {\em commit} primitive has three phases. The first phase is receive data where a {\em commit} request is made using the diff staged in the snapshot. Stepping through this phase brings us to the extend graph phase, where the version history graph is extended from the HEAD version (632e) to the newly created version (3a27) and the new version is marked as the new HEAD. The last phase of commit, which is yet to be executed, is the garbage collect phase where obsolete versions in the graph (in this case 632e) are cleaned up.


At the bottom, we see three buttons: Step Node, Step All Nodes, and Play. Clicking on Step Node, would allow the garbage collect phase of {\em commit} at the Grouper node to be executed. Clicking on the Step All Nodes, would allow all nodes to execute the next phase of the step that they are paused at, if any. Play allows the user to fast forward the execution up until the next breakpoint condition is hit.

Since the {\em fetch}, and {\em push}  primitives of the dataframe span across multiple GoT Nodes, they are broken up into two sets of operations each: {\em fetch} and {\em respond to fetch}, {\em push}, and {\em respond to push}, to observe the state changes at both the node making the request and the node receiving the request.

\subsection{Debugging Word Frequency Counter}
To debug the mismatch between the expected and observed output of the application, we first put in the condition for the breakpoint as seen in Figure~\ref{fig:start_screen}, and hit the play button. All nodes are executed in debug mode and reroute their primitives through the GCN. At each of these rerouted steps, the GCN observes the states of the dataframe in each Node and executes the breakpoint conditions. 

\begin{figure*}
\centering
\includegraphics[width=\textwidth]{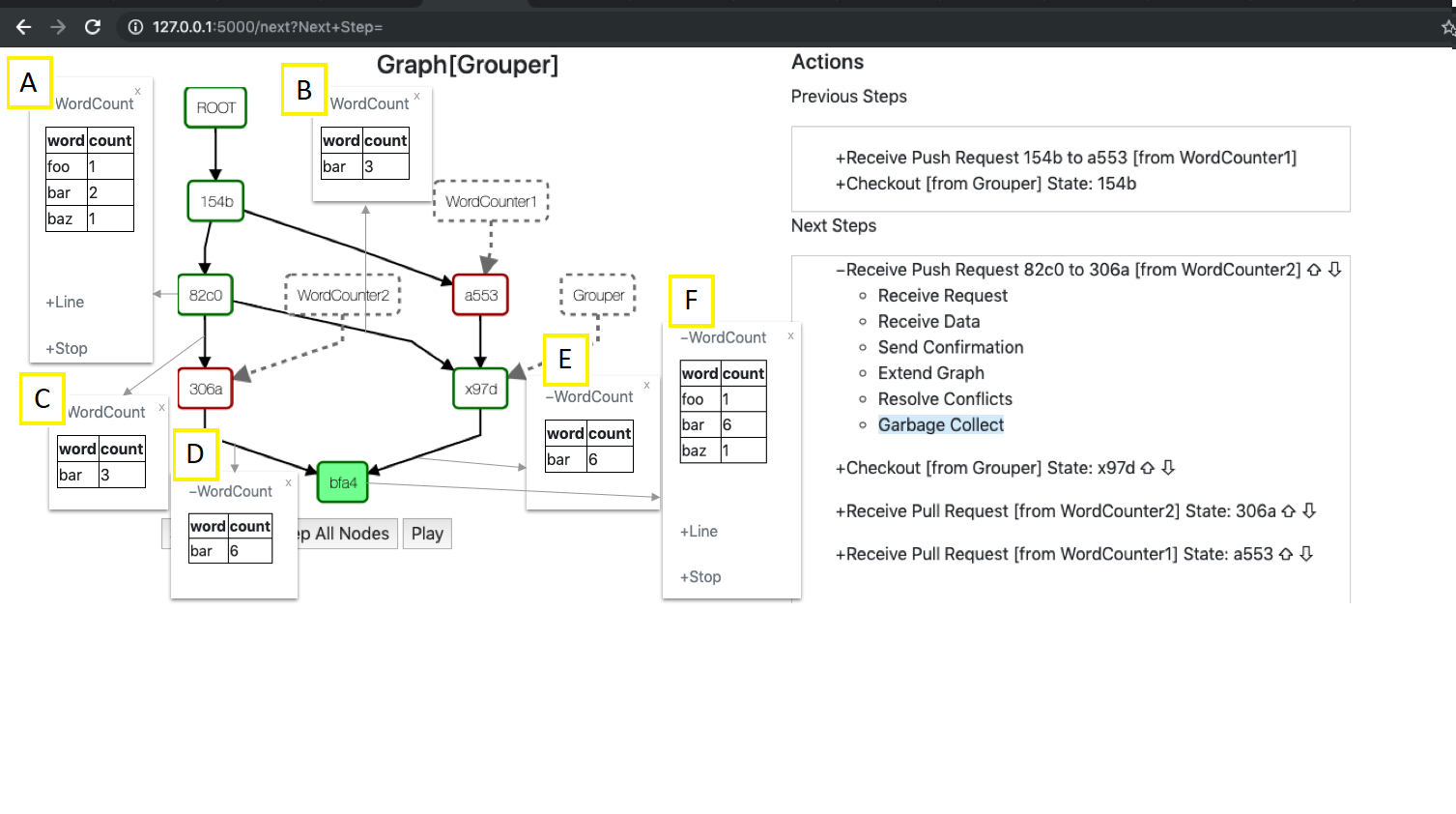}
\caption{Debugger view at Grouper showing response to a {\em push}  request.}
\label{fig:after_push_resp}
\end{figure*}

When the breakpoint is matched, the execution of all nodes is paused and GoTcha shows the view of the Grouper node where the condition matches, shown in Figure~\ref{fig:after_push_resp}. Here, we see that the current step being executed is a {\em push}  request from WordCounter2 from version 82c0 to version 306a. The execution is paused at the start of the garbage collect phase.

The version history contains seven versions. Starting from the top, we have ROOT again as the start version. Version 154b happened-after ROOT. All versions that were present between 154b and ROOT, like the version 632e and 3a27, have all been garbage collected. 

At 154b, we see a fork in the path. Both versions 82c0 and a553 happened-after 154b, and are siblings. These are concurrent updates and were performed on different GoT nodes. Version 82c0 is bordered green while a553 is bordered in red. This means that update 154b $\rightarrow$ a553 was received by the version history at Grouper after the update 164b $\rightarrow$ 82c0. The GoT node resolved such conflicts using the custom merge function written in Listing~\ref{lst:merge}. The output of the merge function was a new version x97d. Since x97d happened-after both the concurrent versions, 82c0 and a553, the graph was updated with the happened-after relations and x97d has two in-edges. Additionally, each of these edges are associated with a diff that transforms the previous version to the version at x97d.

Over the course of execution, another concurrent update was performed with the update 82c0 $\rightarrow$ 306a being concurrent with previously resolved conflict. Another conflict resolution is performed using the merge function. A new resolved version bfa4 is created having a happened-after relation with both 306a, and x97d. The version history is updated to show these relations and the version bfa4 is marked as the current HEAD version of the version history at Grouper.

Looking at the dotted line relations, we see that the snapshot at Grouper is at the version x97d. Additionally, the last know versions of WordCounter1 and WordCounter2 are a553, and 306a, respectively.

The state at version bfa4 is shown in the tooltip F. The tooltip shows us three WordCount objects and the WordCount object for the word `bar' has a count of six, showing us why the conditional breakpoint was hit. Looking at the version at the start of the merge, 82c0, in the tooltip A, we see that the count of `bar' is two. The diffs associated with 82c0 $\rightarrow$ 306a (tooltip C) and 82c0 $\rightarrow$ x97d (tooltip B) both update the count of `bar' to three. This means that both WordCounter1 and WordCounter2 had the count of `bar' as two, and observed a `bar' token updating the count concurrently to three. At the end of the merge function, this count is set to six, and can be see in the diffs for both 306a $\rightarrow$ bfa4 (tooltip D) and x97d $\rightarrow$ bfa4 (tooltip E). This means that the error is in the merge function. We can see that the merge function in Listing~\ref{lst:merge}, on detecting a conflicting count, simply adds up the counts. So receiving two counts of three, would result in a count of six. However, the actual increment in each update is actually just one. The right way to merge counts would be to find the total change in count and add it to the original count. We can fix the code as shown in Listing~\ref{lst:correct_merge} and the word counting application gives the right output.

\begin{lstlisting}[language=Python,basicstyle=\small, numbers=left,
label=lst:correct_merge, captionpos=b, caption=Merge function used at the Grouper node.]
def merge(conf_iter,orig_df,
          your_df,their_df):
    your_df.update_not_conflicting(
        their_df)
    for orig,yours,theirs in conf_iter:
        assert isinstance(
            yours,WordCounter)
        yours.count += theirs.count
        if orig: # False if new objects.
            yours.count -= orig.count
    return your_df
\end{lstlisting}

This bug was found quite easily because GoTcha exposes the version history. By looking at the evolution of the version history, even though the error had already occurred, we could see in which type of state change the error occurred in. In this case, we could see that the version state was correct before the merge function, but after the reconciliation of two correct states, the state was wrong, telling us that the error was in the custom merge function. GoTcha exposes bugs in a Spacetime application in the same way a tool viewing git history can help find the commit that caused a bug in the code. Instead of the evolution of the files being looked at, GoTcha looks at the evolution of the state at each node.
\section{GoTcha: Meeting the Fundamental Requirements}
\label{sec:meeting_req}

In Section~\ref{sec:design} we describe, in detail, the fundamental requirements that an interactive debugger must fulfill. To summarize, the debugger must expose to the user all forms of state changes in the application while minimizing the interference in the natural flow of execution. In this section, we discuss how GoTcha meets these fundamental requirements.  

\subsection{Observing State Changes}
There are three forms of state changes present in a distributed system that are relevant to an interactive debugger: state changes at a node due to local execution, transfer of state between nodes, and the reconciliation of the state received via transmission and the local state at each node. Table~\ref{tab:primitive_function} maps the GoT primitives to the type of state change that it facilitates. In what follows, we explain how GoTcha exposes all these types of state changes to the user.

\begin{table}
    \centering
    \begin{tabular}{l|l}
        {\bf Type of state change} & {\bf GoT Primitives} \\
        \hline
        Change in local state & Commit, Checkout \\
        \hline
        Inter-node state transfer & Push, Respond to Fetch \\
        \hline
        Reconciliation of states & Fetch, Respond to Push,\\ & Commit, Checkout \\
    \end{tabular}
    \caption{Mapping the primitives of GoT to the types of State changes}
    \label{tab:primitive_function}
\end{table}

{\bf Observing changes in local state}: In GoT, the "local state" is the snapshot. The snapshot is updated by write operations directly from the local application code. These kinds of state updates can be observed by traditional debuggers. However, as mentioned in Section~\ref{subsubsec:local_exec}, the amount of state changes in a distributed system can overwhelm the user, and a distributed systems debugger should reduce the number of such updates shown. GoTcha does not track the change of state over every line of code at each GoT node. Instead, it tracks the change in the snapshot over consecutive interactions (commit, and {\em checkout}) between the snapshot and the version history. All changes in between these interactions is purely local and grouped together as one update by both GoT and GoTcha.

{\bf Observing the transfer of state}: The local state of a GoT node is transferred to remote nodes in two ways: a {\em push} from the local node to the remote node, or the response by a remote node to a {\em fetch} from the local node. The user can step through these primitives to observe this communication. Specifically, the user can see when such requests are made, and the delta changes that are transferred as a part of these requests.

{\bf Observing reconciliation of multiple states}: When a node receives state changes transferred from a remote node, it needs to reconcile the states changes. As explained in Section~\ref{subsubsec:recon}, reconciliation is a two step process: first, receiving changes from a remote node, then introducing these changes to the state of the local node. GoTcha must expose both steps to allow the user to observe reconciliation correctly. The first step is observed in GoTcha when observing the state changes on receiving deltas either at the end of the {\em fetch}, or when responding to a {\em push}  request. The acceptance of these changes can be observed during the {\em fetch}, response to a {\em push}, {\em commit}, or {\em checkout}. Conflicts are resolved using custom merge functions that are observed by GoTcha. Changes can also be accepted, as is, without conflicts through a {\em checkout}. All ways of receiving delta changes and observing the acceptance of these changes can be observed by GoTcha allowing the user to observe reconciliation of multiple states.

\subsection{Controlling the Flow of Execution}
GoTcha follows the centralized debugger design explained in Section~\ref{subsec:req2}. The central component, GCN, takes control of all GoT primitives that read or modify the version history. This means that even {\em commit} and {\em checkout} primitives, which are normally local operations, are also routed through the GCN. Control over the execution of the changes to the version history is given to the user. The user can reorder and interleave requests that have to be processed and can explore possible execution variations. This would allow the user to observe if, for example, the conflict resolution functions are performing as intended. The user interface for reordering or interleaving execution steps is shown in Figure~\ref{fig:after_push_resp}, where there are additional steps that are pending at the Grouper node. The developer can reorder and interleave these pending operations using the promote and demote arrows shown on the right side next to each step.

Roll backs are an additional and useful tool to explore different state changes without having to restart the entire execution. Since we have the entire history of execution given to us by the version history, we support roll backs to a previous version. When a roll back in performed, the state in the version history is reverted to an older version. It is important to note that the local state and the execution of the application code is not rolled back. This means that state changes observed after roll backs are only meaningful when the application code at each node is stateless and performs the same action iteratively. However, reconciliation can be observed well using roll backs. 

Rolling back the execution state at a node, along with the state of the dataframe, would require that we either take control of the programming language runtime in each node, which suffers the same problems of coordinating distributed control as discussed in Section~\ref{subsec:req2}, or we integrate GoTcha with a traditional single-threaded debugger at each GoT node. While the first is unfeasible, the second can be a future possibility and is discussed in the next section.

\section{Beyond Interactivity}
\label{sec:discussion}

GoTcha is a good first step into the interactive debugging of distributed applications. By relying on the idea of version control of objects, and exposing the version history at every node, we meet the minimum requirements for observing all forms of state changes in distributed applications. GoTcha fills a hole -- interactivity -- in the tools available for debugging distributed applications. Interactivity has been an elusive piece in this ecosystem, and not much is known about how it can be used in a distributed context. With GoTcha, we see both potential and challenges in the future development of interactive debuggers for distributed applications. Moreover, we envision the development of powerful tools by combining GoTcha with existing concepts related to distributed debugging. In this section, we expand on the potential and challenges of this vision.

\subsection{Scalability of Interactivity}
An inherent property of traditional interactive debuggers is their reliance on the user to explore the possible paths of failures. This is no less true for GoTcha. However, in large systems with large number of possibilities to observe, this interactivity can potentially be overwhelming to the user. Interactive debuggers for single threaded systems ignore this complexity by design, hoping that the advantages offered by the live exploration of the execution of code compensates for the disadvantage of not being able to explore every path and fixing all issues. This exploration space is much larger in distributed systems when compared to single threaded systems because there are more types of state changes that have to be considered. Therefore, in GoTcha, the advantages offered by live exploration of the execution heavily depends on the scalability of both the debugging system, and the user interface, when increasing the number of GoT nodes in the application being debugged.

\subsubsection{Scaling the Debugging System}
In an application with a few number of nodes, the exploration of the execution can be easily visualized and followed and the centralized approach of GoTcha does not hinder the debugging process. However, in applications with a large number of nodes, which is common in a distributed setting, the centralized approach can be a bottleneck. Since every primitive of the dataframe involved in reading or writing to the version history has to be rerouted through the GCN, execution of the application through the debugger is much slower. It would also take longer to hit the conditional breakpoints potentially making the whole debugging process tedious. An easy solution, and one that works with GoTcha as it is, would be to reduce the number of GoT nodes in the application during debugging. The user could debug issues in this much smaller application, fix the problems, and then scale the number of nodes back up. However, this approach might not be always possible and, therefore, changes to the debugging architecture might be needed to solve this problem.

To understand the difficulties involved, let us look at the flow of interactive debugging. There are essentially two modes that GoTcha executes in. First, a ``free-run'' mode, where the application executes as it would under normal conditions until it hits a breakpoint. Second, we have the slow and more deliberate ``step-by-step'' mode which activates when the free-run mode matches a breakpoint, or if the user is exploring execution paths. In the second mode, the user has control over the execution and is observing a small and very specific part of the entire application. The design of GoTcha is tailored towards the step-by-step mode. The central GCN helps coordinate these steps, and visualizations are created with this mode in mind. However, the same central GCN which enables total control during the step-by-step mode, is a bottleneck in the free-run mode when the number of GoT nodes becomes too large. A distributed approach to debugging would be as scalable as the distributed application it is debugging, but only during the free-run mode. However, such an approach would again have a hard time scaling with the number of nodes when the debugger has to control the application in the step-by-step mode. This incompatibility of designs and the multiple modes that interactive debuggers work in, is the underlying reason why the advantages of interactivity in debugging distributed systems diminish with scale.

A possible solution is to change the architecture of GoTcha in each of the modes, matching the strength of each design with the mode that they work best with. In the free-run mode, nodes could communicate directly with each other, and log their activity with the GCN. Each node also receives the list of breakpoint conditions. When a break point is hit, the GCN receives this information and then instructs every other node to switch to the step-by-step mode, taking control of the application and handing it over to the user. 


\subsubsection{Scaling the User Interface}
With a large number of nodes, we also encounter the problem of the user interface having too much information. The network topology is certainly going to be difficult to read making it difficult for the user to take a deep dive into the GoT nodes and explore specific execution paths. Conditional breakpoints become the only way to explore the execution meaningfully making the debugger strictly for finding bugs whose symptoms are already known. A possible solution is to use grouping algorithms to show the topology of the application concisely. Alternately, algorithms like PageRank can be used to show only nodes that are heavily connected.

The view of the version history at each node can have a lot of information when there is significant interaction with the node. For example, a view of the version history at a Grouper node, working with a thousand WordCounter nodes, could potentially have a thousand steps pending at Grouper and waiting to be stepped through. To enhance the navigation of execution, the user interface could allow users to attach breakpoints to the end of the steps that are pending, allowing the user to skip large batches of steps without necessarily having to artificially promote specific the step that they wish to see, to the top of the pending list. Such a breakpoint would be a close match to the non-conditional breakpoints that exist in traditional debuggers.

\subsection{Integration with Alternate Debugging Concepts}
GoTcha is built as a stand alone system, over the GoT model, that helps debug the application by exposing the changes to the version history at each node. State changes within the application are observed from a version control point of view and is observed in broad strokes over several lines of code. The bugs to be found are, however, in these lines of code and integration with traditional interactive debuggers can help find these bugs. As such, GoTcha does not interfere with the use of traditional interactive debuggers at a single node. A single threaded interactive debugger can break down the state changes due to local execution and allow the user to debug the lines of code, while also being sure that the state cannot change in unpredictable ways from one line to the next.

Integration with non interactive forms of distributed debugging are also possible. For example, GoTcha, during the free-run mode, is similar to a tool for record and replay. When a conditional breakpoint is hit, it would be possible for the user to cycle through the previous steps and observe the previous states of the application along with the interactions that occurred between the nodes. Cycling through the previous steps is important because conditional breakpoints are usually used to find the execution point where the symptom of the error manifests. This may not always be the point where the error is. The user can find these errors by observing previous states of the application. If the user does not put a conditional breakpoint and executes the entire application in free-run mode, the entire execution is recorded and can be replayed. Existing tools and research on record and replay can add value to the free-run mode of GoTcha and make it a more powerful tool. The integration of non interactive debugging tools would enhance the approach of finding errors during free-run as most of these tools deal with postmortem analysis of execution, while the interactivity of GoTcha would allow the user to observe errors during the exploration of live execution of the application.

\section{Conclusion}
\label{sec:conclusion}

Interactive debuggers for distributed systems is an understudied area of research. In this paper, we discuss the major goals that interactive debuggers for a distributed systems should meet. In addition to exposing state changes at a node through local processes like traditional interactive debuggers, interactive debuggers for distributed systems should expose the communication between nodes, and the integration of this information that is communicated, at each site. The debuggers, should be able to expose these information exchanges while giving the user complete control over the execution of the system. In order for interactive debuggers to meet these requirements, support is needed from the underlying programming model. We discuss the specific features required from the programming model that can facilitate interactive debuggers. We put our theory to test by describing the implementation of an interactive debugger called GoTcha over the distributed programming model called GoT. We discuss the design of GoT, the design of GoTcha, and describe a simple debugging process using the example of a distributed word frequency counter. GoT, based on distributed version control systems like Git, tracks the change of state at each node in a version history. GoTcha exposes the changes made to the version history to the user, and gives the user complete control over the execution of the read and write functions of the version history. Finally, we discuss the potential and challenges involved in further developing interactive debuggers. We hope that the design of the interactive debugger detailed, and the tool GoTcha, can inspire more research into making interactive debugging feasible.


\bibliography{onward_main}

\appendix
\end{document}